\documentclass[a4paper,twocolumn,showpacs,amsmath,pra,amssymb,nofootinbib]{revtex4-1} 
\usepackage[T1]{fontenc}
\usepackage[latin9]{inputenc} 
\usepackage{times}
\usepackage{color}
\usepackage{xspace}
\usepackage{amssymb,amsmath}
\usepackage{amsbsy}
\usepackage{graphicx}
\usepackage{epstopdf}
\usepackage{float}

\newcommand{\etal}{\textit{et al.}~}

\newcommand{\eqr}[1]{Eq.~(\ref{#1})}
\newcommand{\eqrs}[1]{Eqs.~(\ref{#1})}
\newcommand{\figr}[1]{Fig.~\ref{#1}}
\newcommand{\figsr}[1]{Figs.~\ref{#1}}
\newcommand{\appr}[1]{Appendix \ref{#1}}
\newcommand{\secr}[1]{Sec.~\ref{#1}}
\newcommand{\refr}[1]{Ref.~\cite{#1}}
\newcommand{\refrs}[1]{Refs.~\cite{#1}}

\newcommand{\braca}[1]{\left(#1\right)}
\newcommand{\bracb}[1]{\left[#1\right]}
\newcommand{\bracc}[1]{\left|#1\right|}
\newcommand{\bracd}[1]{\left\{#1\right\}}

\newcommand{\bp}{\mathbf{p}}
\newcommand{\br}{\mathbf{r}}
\newcommand{\bq}{\mathbf{q}}

\newcommand{\df}{f\braca{\bp,\br,t}}
\newcommand{\denr}{n\braca{\br,t}}

\newcommand{\gradr}{\nabla_{\br}}
\newcommand{\gradp}{\nabla_{\bp}}
\newcommand{\potr}{U\braca{\br,t}}
\newcommand{\icol}{I\bracb{f}}
\newcommand{\dcs}{\frac{d\sigma}{d\Omega}}
\newcommand{\incoming}{f'f'_{1}}
\newcommand{\outgoing}{ff_{1}}

\newcommand{\phasel}{\delta_{l}}
\newcommand{\lpoly}{P_{l}\braca{\cos \theta}}

\newcommand{\Kn}{\mathrm{K}\!\mathrm{n}}

\newcommand{\np}{\mathcal{N}_{P}}

\newcommand{\nt}{\mathcal{N}_{T}}
\newcommand{\nsc}{\mathcal{N}_{sc}}

\newcommand{\ts}{\Delta t}

\newcommand{\nth}{N_{th}}

\newcommand{\tcs}{\sigma \braca{v_r}}
\newcommand{\cellv}{\Delta V_c}

\newcommand{\pcol}{P_{ij}}
\newcommand{\pcolr}{\tilde{P}_{ij}}
\newcommand{\mr}{\tilde{M}_{c}}

\newcommand{\mkrcs}{\bracb{ v_r \tcs}_{\max}}

\newcommand{\tsg}{\delta t_i}
\newcommand{\tsc}{\delta t_c}

\newcommand{\dfeq}{f_{eq}\braca{\bp,\br}}
\newcommand{\pot}{U(\mathbf{r})}

\newcommand{\omegaz}{\omega_{z}}

\newcommand{\dfcol}{f_{coll} \braca{\bp,\br}}
\newcommand{\erf}[1]{\mathrm{erf}\braca{#1}}

\newcommand{\phasea}{\delta_{0}}
\newcommand{\phaseb}{\delta_{2}}
\newcommand{\ucos}{\braca{3\cos^{2}\theta - 1}}

\newcommand{\tcoll}{T_{coll}}

\begin{document}

\title{Direct Simulation Monte Carlo method for cold atom dynamics: \\
 classical Boltzmann equation in the quantum collision regime}
\author{A.~C.~J.~Wade} 
\author{D.~Baillie} 
\author{P.~B.~Blakie}  

\affiliation{Department of Physics, Jack Dodd Centre for Quantum Technology, University of Otago, Dunedin, New Zealand.}

\begin{abstract}
In this paper we develop a direct simulation Monte Carlo (DSMC) method for simulating highly nonequilibrium dynamics of nondegenerate ultra cold gases.
We show that our method can  simulate the high-energy collision of two thermal clouds in the regime observed in experiments [Thomas \etal Phys.~Rev.~Lett.~\textbf{93}, 173201 (2004)], which requires the inclusion of beyond $s$-wave scattering. We also consider the long-time dynamics of this system, demonstrating that this would be a practical experimental scenario for testing the Boltzmann equation and studying rethermalization. 
\end{abstract}


\maketitle


\section{Introduction}
 
Within ultra-cold-atom research, there are a range of problems requiring the understanding of the dynamics of a normal gas. For example, studies of collective modes of Bose \cite{Shvarchuck2003a}  and Fermi \cite{Kinast2004a} gases (also see \refrs{Jin1996a,Stamper-Kurn1998a}), spin waves \cite{Lewandowski2002a,McGuirk2002a}, hydrodynamic expansion of a Bose gas near the critical temperature  \cite{Gerbier2004a}, and more recently, the dynamics and thermalization of a nearly degenerate gas of polar molecules \cite{Ni2010a}.
 These are all regimes in which the Boltzmann equation is thought to provide an accurate description. In many of these cases, the system is only weakly disturbed from equilibrium, and some approximate solution can be provided using a relaxation approximation for the collision integral and some form of linearization \cite{Nikuni2002a}, scaling \cite{Al-Khawaja2000a,Pedri2003a}, or variational \cite{Guery-Odelin1999a} ansatz.   For more strongly dynamical situations, these approaches are insufficient, however, the direct solution of the Boltzmann equation for the six-dimensional distribution function is generally considered intractable and is normally tackled using some form of stochastic particle simulation. Some applications of such calculations include the work of Wu and co-workers \cite{Wu1996a,Wu1997a,Wu1998a} on evaporative cooling and expansion dynamics, Jackson and co-workers \cite{Jackson2001a,Jackson2001b,Jackson2002a,Jackson2002b,Jackson2002c} on bosonic collective-mode dynamics (coupled to a superfluid by the Zaremba-Nikuni-Griffin (ZNG) formalism \cite{Zaremba1999a}), the work of Urban and Schuck \cite{Urban2006a}, Urban \cite{Urban:2007a,Urban:2008a}, and Lepers \etal \cite{Lepers2010a} in formulating fermion dynamics (see also \refrs{Vignolo2002a,Toschi2003a,Capuzzi2004a, Toschi2004a}), and Barletta \etal \cite{Barletta2010a} and Barletta \cite{Barletta2011a} in describing sympathetically cooled molecular gases.  
\begin{figure}[h]
\begin{center}
    \includegraphics[width=.8\linewidth]{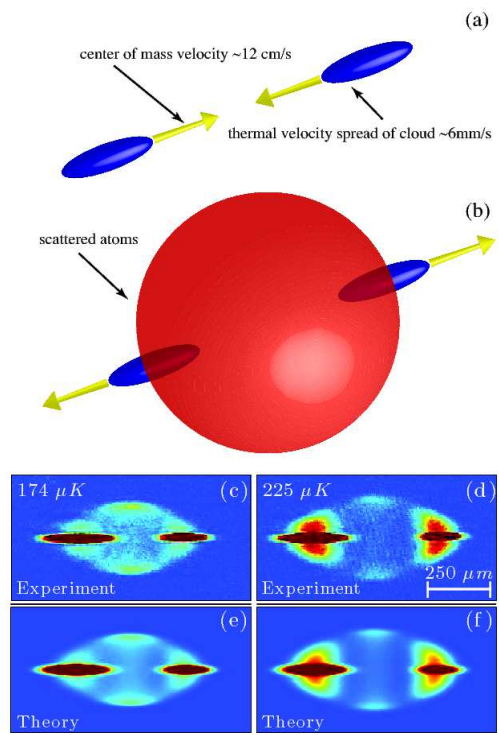}
\caption{(Color online) Ultra cold atom collider: (a) schematic of the precollision arrangement of two clouds at $\sim200$ nK approaching at a collision energy of $\sim200\,\mu$K; (b)  schematic of a postcollision system. (c) and (d) experimental images of post scattering density for two collision energies spanning the $d$-wave shape resonance. (e) and (f) show the theoretical calculations matching the experimental results using the direct simulation Monte Carlo (DSMC) method developed in this paper.\label{fig:colliderimage}}
 \end{center}
\end{figure}

Here, we develop an algorithm for simulating the Boltzmann equation that is significantly more accurate and efficient than these previous methods and is applicable to more extreme regimes of dynamics. Indeed, our main motivation was to develop a theory capable of describing the ultra-cold-atom collider developed by the Otago  group \cite{Thomas:2004a,Kjaergaard2004a,Mellish2007a}. In those experiments  (nonquantum degenerate), clouds of bosonic atoms at a temperature of $\sim 200$ nK 
were accelerated and were collided at an energy of $\sim200$ $\mu$K (see Fig.~\ref{fig:colliderimage}). Several features of these experiments make the numerical simulation difficult: 
\begin{itemize}
\item[(i)] The system is far from equilibrium and accesses a large volume of phase space. A good representation of each cloud before the collision requires nano-Kelvin energy resolution, however, during the collision, atoms are scattered over states on the collision sphere with an energy spread on the order of a milli-Kelvin.
\item[(ii)] The collision energies are sufficiently large that an appreciable amount of higher-order (i.e. beyond s-wave) scattering occurs. In particular, in  experiments $p$-wave scattering \cite{Mellish2007a} and a $d$-wave \cite{Thomas:2004a} shape resonance have been explored.
\end{itemize}
The algorithm we develop is suitable for this regime, and, as shown in  Figs.~\ref{fig:colliderimage}(c)-\ref{fig:colliderimage}(f), it can provide a quantitative model for the experimental data in Ref.~\cite{Thomas:2004a}. Feature (i) discussed above presents a great challenge, and using the traditional Boltzmann techniques  employed to date in ultra-cold-atom research, this would require super computer resources. We show how to make use of an adaptive algorithm (that adapts both the spatial grid and the times steps to place resources where needed)  to accurately simulate an ultra cold collider on commodity personal computer hardware.

We note that, in addition to collider experiments, a capable Boltzmann solver would allow theoretical studies in a range of areas of emerging interest, such as the turbulence and flow instabilities in the normal phase of a quantum gas. Here, we will focus on the classical regime where the phase-space density is small compared to unity such that the many-body effects of Bose-stimulated or Pauli-blocked scatterings are negligible. However, the systems we consider will be in the \textit{quantum collision regime}, whereby the thermal de Broglie wavelength is larger than the typical range of the interatomic potential. Notably, in this regime, the scattering is wave like, and quantum statistics on the two-body level gives rise to profound effects in the individual collision processes, even though many-body quantum statistics is unimportant.

All of the Boltzmann simulations appearing in the ultra-cold-atom literature have been based on DSMC-like methods, typically employing the algorithm described in Bird's 1994 monograph \cite{Bird:1994a}. 
However, a challenging feature of ultra cold gases is that the local properties (e.g., the density) can vary by orders of magnitude across the system, and no single global choice of parameters for the DSMC can provide a good description across this entire range. For this reason, we introduce the use of two locally adaptive schemes to allow the system to refine the description and to allocate more computational resources  to regions of high density. These schemes, which we discuss in Sec.~\ref{SecDSMC}, are as follows: locally adaptive time steps (LATSs) and locally adaptive cells (LACs).

In Sec.~\ref{SecTests}, we validate our algorithm using a variety of tests to demonstrate its applicability and performance. Then, in Sec.~\ref{SecCollider}, we apply it to the regime of the ultra-cold-atom collider experiments \cite{Thomas:2004a}.

\section{Theory}

\subsection{Boltzmann equation} \label{sec:QBE}

The system is described semiclassically by the phase space distribution function 
$f\equiv\df$, which evolves according to the  Boltzmann equation \cite{Pethick:2001a},
\begin{equation}
\bracb{\frac{\partial}{\partial t}+\frac{\bp}{m}\cdot \gradr-\frac{1}{m}\gradr \potr \cdot \gradp }f=\icol,\label{eq:BE}
\end{equation}
and the position space density of the atoms $\denr$ is given by
\begin{equation}
\denr=\int\frac{d^{3}p}{h^3}\df.
\end{equation}
The left-hand side of \eqr{eq:BE} describes the evolution of atoms under the potential $\potr$. In general, $\potr$ may contain a mean-field term, however, for our analysis in this paper, we only consider the case where $\potr$ is an external trapping potential.

The collision integral $\icol$, accounts for the collisions between atoms and is given by
\begin{align}
&\icol=\frac{1}{m} \int\frac{d^{3}p_{1}}{h^3}\int d\Omega \dcs  \bracc{\bp_1-\bp} \left[\incoming
-\outgoing\right],\label{eq:icol}
\end{align}
where $\dcs$ is the differential cross section and $f_{1}\equiv f\braca{\bp_{1},\br,t}$,  $f'\equiv f\braca{\bp',\br,t}$, etc. When considering the flow of atoms through phase space due to collisions, $\icol$ has a simple interpretation. The term in \eqr{eq:icol} containing $ff_{1}$ describes collision events where the atoms are initially at the phase-space points $\braca{\bp,\br}$ and $\braca{\bp_{1},\br}$ and have final states $\braca{\bp',\br}$ and $\braca{\bp'_{1},\br}$. The rate of such a collision depends on the densities of the initial states $ff_{1}$, kinetic factors described by the differential cross section, and the flux of incident particles, which is proportional to $\bracc{\bp_1-\bp}$. The opposite process where atoms scatter to $\braca{\bp,\br}$ and $\braca{\bp_{1},\br}$ is accounted for by the $f'f'_{1}$ term. 
The quantum statistics of the atoms can be included by the addition of  $\braca{1\pm f'}\braca{1\pm f'_{1}}$ terms in the collision integral, which account for  Bose-stimulated scattering ($+$) or Pauli blocking ($-$). Here, we neglect quantum statistics as is appropriate for the regime where $f\ll1$ and will address considerations for the full quantum Boltzmann equation elsewhere \cite{Wade:2010a}.

\subsection{Partial-wave treatment of collisions}
While our interest here is in ultra cold gases with sufficiently low phase-space density to neglect many-body quantum statistics, the two-body collisions themselves are in the quantum collision regime and are conveniently characterized in terms of a partial-wave expansion. The differential cross section for identical bosons ($+$) or fermions ($-$) in the same internal state  is
\begin{equation}
\dcs =  \bracc{f_{sc}\braca{\theta}\pm f_{sc}\braca{\pi-\theta}}^{2}, \label{eq:dcs}
\end{equation} 
where
\begin{equation}
f_{sc}\braca{\theta} =  \frac{\hbar}{i m v_r }\sum_{l=0}^{\infty}\braca{2l+1}\braca{e^{2i\phasel}-1}\lpoly , \label{eq:fsc}
\end{equation}
is the scattering function, $v_r$ is the magnitude of the relative velocity of the colliding particles, $\phasel$ is the phase shift associated with partial wave $l$, $\lpoly$ is the $l^{\mathrm{th}}$ Legendre polynomial, and $\theta$ is the centre-of-mass scattering angle. In general, the phase shifts have a collision energy dependence ($v_r$), which is a nontrivial task to calculate.

For bosons (fermions), the total wave function is required to be symmetric (antisymmetric),  and hence, only the even (odd) $l$ terms in \eqr{eq:fsc} contribute to the differential cross section. In this paper, we focus on the case of bosons, motivated by the experimental work we seek to describe  \cite{Thomas:2004a}.

\section{DSMC method}\label{SecDSMC}

\subsection{Background for DSMC}

The DSMC method is the most widely used tool for modeling fluid flow on the subcontinuum scale and has found itself successfully applied to a huge range of physics from shock waves \cite{Bird:1994a} and Rayleigh-B\'{e}nard flow \cite{Watanabe1994a} to aerodynamics of spacecraft \cite{Oran1998a}, chemical reactions \cite{Anderson2003a}, microfluidics \cite{Frangi:2008a}, acoustics on Earth, Mars, and Titan \cite{Hanford2009a}, volcanic plumes on Jupiter's moon Io \cite{Zhang2004a}, and much more.

These situations are characterized by being dilute (two-body collisions) and having a high Knudsen number ($\Kn$), which is given by the ratio of the mean-free path to the representative length scale of the system.  For $\Kn\gtrsim0.1$, a microscopic kinetic theory is necessary, while for $\Kn\lesssim0.1$, the system tends to be sufficiently hydrodynamic for a continuum approach to be applicable for understanding coarse-grained dynamics.\footnote{In the cold-atom community, it is more common to specify these regimes as $\omega\tau$, where $\omega$ is the excitation frequency and $\tau$ is the collision time.} This is not to say that DSMC is inapplicable or is inefficient in this regime; indeed, recently, Bird has shown that, in nonequilibrium situations with $\Kn\sim0.01$, the DSMC algorithm (employing many of the techniques we introduce for cold atoms here) can be more accurate and efficient than Navier-Stokes methods, while also providing details of the microscopic (subcontinuum) dynamics \cite{Bird2007a}. We also note that the \textit{consistent Boltzmann algorithm} \cite{Alexander:1995a} was developed by making an adjustment to the DSMC algorithm, where the positional shifts associated with collisions are taken into account, giving the correct hard-sphere virial. This allows for exploration into even lower Kn and has been explored in the context of quantum nuclear flows \cite{Kortemeyer:1996a,Morawetz:1999a}.

For reference, cold-atom experiments often operate in the collisionless regime ($\Kn>1$), however, values of $\Kn\sim 0.01$ have been explored, e.g., the above-critical temperature collective modes of a $^{23}$Na gas studied by Stamper-Kurn \etal \cite{Stamper-Kurn1998a} had $\Kn\sim0.1$;  Shvarchuck \etal  \cite{Shvarchuck2003a}  studied the hydrodynamical behavior of a normal $^{87}$Rb gas in which $\Kn\sim0.02-0.5$.

\subsection{Overview of formalism and general considerations}
In the DSMC method, the distribution function is represented by a swarm of test particles,
\begin{equation}
\df \approx  \alpha \, h^3\sum_{i=1}^{\nt} \delta \bracb{\bp-\bp_{i}\braca{t}}  \delta \bracb{\br-\br_{i}\braca{t}} , \label{eqn:df}
\end{equation}
where $\alpha=\np/\nt$ is the ratio of physical atoms ($\np$) to test particles ($\nt$). 
These test particles are evolved in time in such a manner that $\df$ evolves according to the Boltzmann equation. 

The basic assumption of DSMC is that the motion of atoms can be decoupled from collisions on time scales much smaller than the mean-collision time. In practice, this means that a simulation is split up into discrete time steps $\ts$, during which, the test particles undergo a collisionless evolution, then collisions between test particles are calculated. 

The relation of the test particles to physical atoms is apparent in Eq.~(\ref{eqn:df}) when $\alpha=1$, but, in general, they are simply a computational device for solving the Boltzmann  equation. In many conventional applications of DSMC, good accuracy can be obtained with $\alpha\gg1$ (i.e., each is a \textit{super particle} representing a larger number of physical atoms), however, in our applications on nonequilibrium dynamics of ultra cold gases, we often require $\alpha\ll1$. Increasing the number of particles improves both the accuracy and the statistics of the simulation, and in highly nonequilibrium situations, it can be essential to have a large number of particles. The DSMC method is designed so that the number of computational operations per time step scales linearly with the number of particles, i.e., $O\braca{\nt}$. The recent work of Lepers \etal \cite{Lepers2010a} departs from DSMC by using a stochastic particle method similar to that developed in nuclear physics for the simulation of heavy-ion collisions \cite{Bertsch1988a,Bonasera1994a}, which tests if two particles are at their closest approach in the present time step, causing the algorithm to scale as $O\braca{\nt^2}$. These methods have been reformulated in terms of DSMC by Lang \etal \cite{Lang1993a}. We typically use $\nt=10^5-10^7$ test particles, and, by the various improvements we describe below, in most cases considered, here, we can obtain accuracy to within $1\%$.

As pointed out in \secr{sec:QBE}, the Boltzmann equation has a simple interpretation in terms of the flow of atoms through phase space. Hence, the collisionless evolution of the test particles is performed by solving Newton's laws for the potential $\potr$, and collisions are governed by the collision integral [\eqr{eq:icol}]. The collisions are implemented probabilistically 
(see \secr{sec:colprob}) using a scheme that requires the particles to be binned into a grid of cells in position space. This serves two purposes: (i) It allows for the sampling of the distribution function, and (ii) it establishes a computationally convenient mechanism for determining which particles are in close proximity. Thus, the accuracy of DSMC depends on the discretization of the problem, the cell size, the time step, and $\nt$. It has been shown to converge to the exact solution of the classical Boltzmann equation in the limit of infinite test particles, vanishing cell size, and vanishing time step \cite{Wagner:1992a}. 

In the original DSMC algorithm \cite{Bird:1994a}, a test particle may collide with any other particle within the cell. This coarse grains position and momentum correlations, such as vorticity, to be the length scale of the cells, as observed by Meiburg \cite{Meiburg:1986a}. If the cells are not small enough, this transfer of information across a cell could lead to nonphysical behavior. To combat this, we have employed a nearest-neighbor collision scheme \cite{Gallis:2009a} outlined in \secr{sec:NNC}, where the collision partner of a particle must be chosen from the nearest neighbors. Although nearest-neighbor collisions alleviate this problem, the cell sizes still must be small in comparison to the local mean-free path 
and the length scale over which the density varies for accurate sampling. 

The time step of the simulation must also be small in comparison to the smallest local mean-collision time to ensure the validity of the basic assumption of DSMC and that physical atoms do not propagate further than the local mean free path before colliding. To ensure this (and for added efficiency), we implement locally adaptive time steps \cite{Gallis:2009a} where, instead of a single global time step, the time step can vary over the whole system, adapting to the local environment.

\subsection{Implementation of DSMC}

Here, we consider the basic implementation of DSMC; a collisionless evolution followed by a collision step where test particles are binned in position space and collisions between them are implemented stochastically via a collision probability. We also discuss the various adaptive schemes we employ for better accuracy and efficiency, while retaining the desired linear scaling of the computational complexity with test particle number.

\subsubsection{Collisionless evolution} \label{sec:evol}
The collisionless evolution is performed by a second-order symplectic integrator \cite{Yoshida:1993a,Jackson2002a}, which updates the phase-space variables of the $i^{\mathrm{th}}$ test particle in three steps:
\begin{subequations}
\begin{gather}
\bq_{i} = \br_{i}\braca{t} + \frac{\ts}{2 m} \bp_{i}\braca{t},  \label{collevolv1}\\
\bp_{i}\braca{t+\ts}  = \bp_{i}\braca{t} - \ts \, \nabla_{\bq_{i}} U \braca{\bq_{i},t},  \\
\br_{i}\braca{t+\ts} =\bq_{i}  +\frac{\ts}{2 m}\bp_{i}\braca{t+\ts}.\label{collevolv3}
\end{gather}
\end{subequations}
Symplectic integrators have the properties of conserving energy and phase-space volume over long periods of time.

\begin{figure}[h]
\begin{center}
    \includegraphics[width=.9\linewidth]{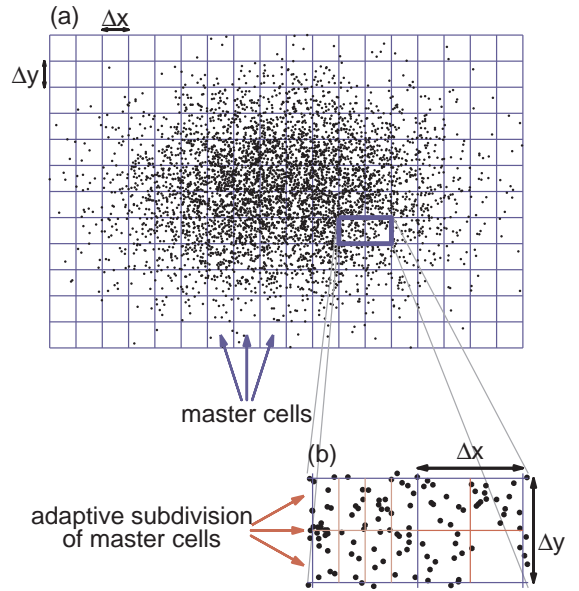}
\caption{(Color online) A two-dimensional schematic of the cells used for a swarm of test particles.
(a) The rectangular master cells are all of the same size and are chosen to ensure all particles lie within the boundaries of this grid. Cell boundaries are indicated by lines, and particles are indicated by dots.  (b) An enlargement of two master cells showing their adaptive subdivisions into smaller cells.  The number of subcells is determined by the number of particles within the master cell. \label{fig:diagr1}}
 \end{center}
\end{figure}

\subsubsection{Master grid and LACs} \label{sec:acells}

To perform collisions, we must first bin the test particles into a grid of cells according to their position. Collision partners are then selected from within each cell. In general, the binning occurs in up to two levels: (i) the \textit{master grid} on which each master cell is a rectangular cuboid of equal size [see Fig.~\ref{fig:diagr1}(a)] and (ii) the adaptive subdivision of the master cells into smaller cells dependent on the number of particles in the parent master cell, i.e., LACs [see Fig.~\ref{fig:diagr1}(b)], which is an optional refinement. The use of several LAC schemes in DSMC is discussed in \refr{Bird:1994a}. It is a useful refinement to the algorithm for applications to cold-atom systems because these typically have large variations in density (such a scheme has been employed in \refr{GueryOdelin1998a} to account for the large change in density during evaporative cooling of a cloud of cesium atoms).
We now discuss these levels in further detail.

At the beginning of the collision step, the grid of master cells is chosen to ensure all particles are held within its boundaries  [see Fig.~\ref{fig:diagr1}(a)]. We choose to keep the size of the master cells in each direction constant in time so that if the particles spread out further in space during the simulation, we add extra cells rather than changing the size of the cells. The particles are then binned into these master cells, and the number of particles in each cell $N_c$ is stored.

For adaptive subdivision, each master cell is considered in turn, and the particles are binned further into a grid of smaller subcells according to $N_c$ [see Fig.~\ref{fig:diagr1}(b)]. Because the number of collisions within a cell increases with density (i.e., number of particles), the subdivision of highly occupied master cells gives a finer resolution of spatial regions where the local collision rate is highest and, hence, more accurate simulations.

Our subdivision procedure aims to produce cells in which the average number of particles is close to some threshold value $\nth$ for which the choice of is discussed in \secr{sec:Accuracy}.  In our algorithm, we do this by finding the integer $l$ such that  $N_c/2^l$ is closest to, but not less than, $\nth$. The master cell is then subdivided into $2^l$ subcells, while giving no preference to any direction in this subdivision. We choose this division scheme over more complicated schemes, as when additionally implementing LATS, the protocol for dynamically changing grids becomes simpler.

We have adopted the notation of specifying quantities pertaining to a particular cell by a subscript $c$. In what follows, when referring to cells, we will mean finest level of cells, i.e., the subcells (if used) or master cells otherwise (e.g., if LAC is being used, $N_c$ refers to that number of atoms in the subcell). We do not explicitly label the cells, indeed, this is to partly emphasize that the calculations performed in each cell are independent of other cells. Thus, the algorithm is intrinsically parallel and is suitable for implementation on parallel platforms  (e.g., see \refr{Frangi:2008a}).

\subsubsection{Collisions: Scaling}\label{sec:colprob}

The collision probability for a pair of test particles $i$ and $j$ in a cell of volume $\cellv$ is given by 
\begin{equation}
\pcol=\alpha\frac{\ts}{\cellv} v_r \tcs, \label{eq:colp1}
\end{equation}
where $\tcs$ is the total cross section. 
This collision probability can be derived from the collision integral (\ref{eq:icol}) via the Monte Carlo integration \cite{Jackson2002a,Wade:2010a}, the kinetic arguments \cite{Bird:1994a}, or the elementary scattering theory \cite{Bonasera1994a}.
The correct collision rate is established by testing  $M_{c}=N_c\braca{N_c-1}/2$ collisions in the cell (see \appr{app:M_c} for justification of this choice). This is inefficient as the number of operations scales as $\nt^2$, and the collision probability may be far less than 1. However, within a cell, the collision probabilities and the number of tested collisions can be rescaled by a single parameter $\Lambda$ such that the number of operations scales as $\nt$ \cite{Bird:1994a},
\begin{subequations}
\begin{gather}
\pcol\to \pcolr=\frac{\pcol}{\Lambda},\label{pscale}\\
M_c\to \mr= {M_c}{\Lambda},\label{Mscale}
\end{gather}
\end{subequations}
and still converge to the same Boltzmann equation evolution. Here, $\Lambda$ is chosen to be
\begin{equation}
\Lambda= \frac{\left\lceil M_c \alpha\frac{\ts}{\cellv}\mkrcs \right\rceil}{M_c}, \label{Lambda1}
\end{equation}
where  $\mkrcs$ is the maximum of this quantity over all pairs of particles in the cell and $\lceil x\rceil$ denotes the ceiling function. This corresponds to Bird's proposal of using $\Lambda=\max\bracd{\pcol}$ \cite{Bird:1994a}, while we ensure that $M_c$ is an integer and at least one collision is tested (\figr{fig:derror} demonstrates the reduction in collisions if this is not taken into account). With this choice of scaling, the maximum collision probability within the cell is $\le 1$ (expected to be close to $1$), and the number of collisions that need to be tested is reduced to  
\begin{equation}
\mr = \left\lceil \frac{N_c-1}{2} n_c \ts\mkrcs \right\rceil, \label{eq:numcol}
\end{equation}
where 
\begin{equation}n_c=\alpha N_c/\cellv,\label{nc}
\end{equation} is the density in the cell.

This enhancement of efficiency is often missed by other stochastic particle methods, or the collisions are adjusted in some other manner. 
For example, Tosi \etal \cite{Toschi2003a} introduced a scheme for fermions where collision pairs with small classical collision probability were neglected.

\subsubsection{Collisions: Nearest-neighbor selection of partners}  \label{sec:NNC}

We employ a nearest neighbor collision scheme to combat discretization effects from finite cell sizes, in particular, the so-called \textit{transient adaptive subcell} (TASC) scheme \cite{Gallis:2009a}. 
\begin{figure}[h]
\begin{center}
    \includegraphics[width=.65\linewidth]{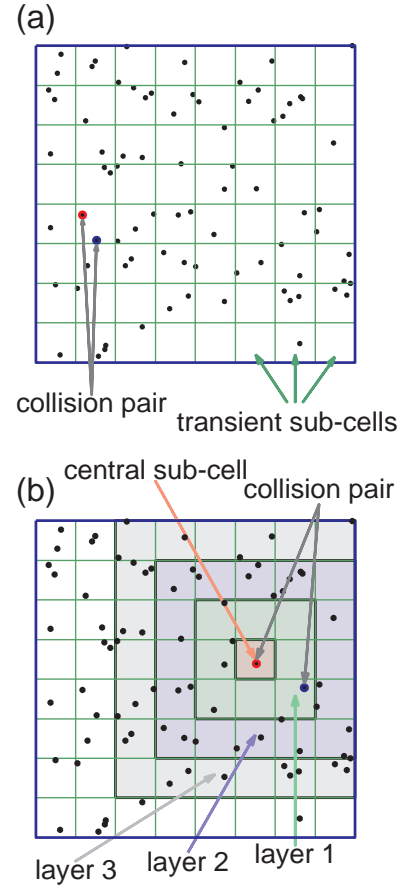}
\caption{
(Color online) A two-dimensional schematic of how collisions are performed within the TASC scheme. 
A single cell (outer boundary line) and the distribution of test  particles (black dots)  are shown in (a) and (b) for two different random collisions. The finer grid of internal lines represents the boundaries of the TASC subcells. The first particle of the collision pair is selected at random from all the particles in the cell. 
In (a), the first particle occupies a TASC subcell that contains other particles, and the second participant in the collision   is chosen at random from these other particles. In (b), the first particle (which occupies  the \textit{central subcell}) is the sole occupant of a TASC subcell. In this case, we check to see if there are any particles in layer 1, and if so,  the collision partner  is chosen at random from these other particles. If there were no particles in layer 1, we would then check layer 2, and so on. \label{fig:diagr2}}
 \end{center}
\end{figure}
Simple sorting of the test particles for the nearest neighbors scales quadratically with the particle number. 
The TASC sorting scheme retains linear scaling, but it does not guarantee the exact nearest neighbor.

The basic TASC scheme is to further bin the particles into subcells within the cell [see Figs.~\ref{fig:diagr2}(a) and \ref{fig:diagr2}(b)], the number of which is roughly equal to $N_c$. In our case, the number of subcells in each direction is equal and is given by $\lfloor \sqrt[3]{N_c} \rfloor$ (with $\lfloor x\rfloor$ as the floor function). When a particle is randomly picked for a collision, its collision partner is established by looking within the immediate TASC subcell [Fig.~\ref{fig:diagr2}(a)], and if not found [Fig.~\ref{fig:diagr2}(b)], each layer starting closest to the particle is searched for other particles. If a layer contains more than one particle, the collision partner is randomly chosen from that set to avoid any biasing. This reduces the distance between colliding pairs significantly and may be decreased even more by increasing $\nt$.

We use this procedure to select each of the $\mr$ pairs of particles for testing if a collision occurs. We also ensure a particle does not undergo a second collision in the same time step.

\subsubsection{Collisions: Testing and implementation of collisions}  

For each of the pairs, the collision goes ahead if $R<\pcolr$, where $R$ is a random number uniformly distributed between $0$ and $1$. As \eqr{eq:BE} describes binary collisions of point like particles that conserve total energy and momentum, only the momenta are changed by keeping the total momentum constant, and the relative momentum vector is rotated about its center \cite{Huang:1987a}. The scattering angles are determined by using an acceptance-rejection Monte Carlo algorithm for the differential cross section.

\subsubsection{LATSs} \label{sec:LATSs}

\begin{figure}[h]
\begin{center}
    \includegraphics[width=.95\linewidth]{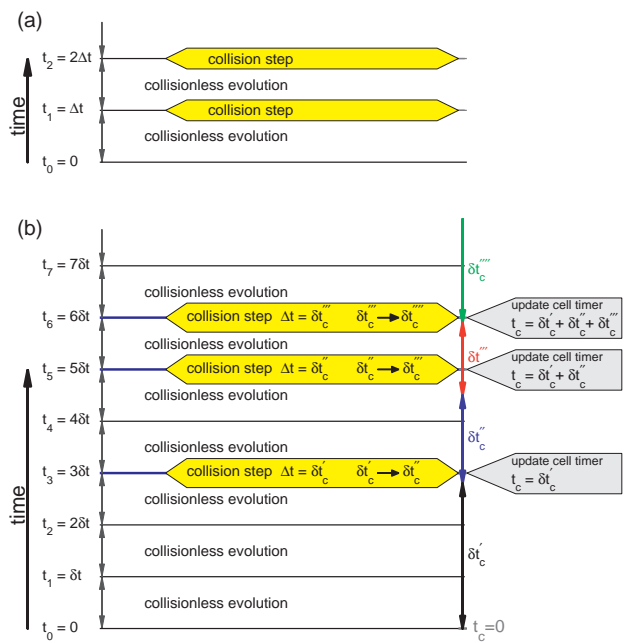}
\caption{
(Color online) An example of the sequence of steps in a DSMC evolution. (a) A simple DSMC scheme where the whole system evolves according to a single global time step $\Delta t$. (b) An example of a cell using LATS. In this example, the global time step ($\delta t$) is held constant, while the local time step ($\delta t_c$) is shown to vary. Collisionless evolution occurs at each global time step. A collision step is performed at the global step when, at least, $\delta t_c$ has passed since the last collision step. At global time $t_3$, we show a collision step, at which the local time counter ($t_c$) is updated and a new local time step ($\delta t_c''$) is established. Here, the local time step decreases, showing two further collision steps that follow shortly after the first.
 \label{fig:diagr3}}
 \end{center}
\end{figure}

All of the preceding aspects of our implementation of DSMC can be performed with the single global time step $\ts$ for all cells such that the evolution of the system is simulated at the times $t_k=k\,\ts$, with $k$ as an integer. At each of these steps, the collisionless evolution is performed, then, is followed by the collision step [see Fig.~\ref{fig:diagr3}(a)]. However, if there is large variation in the properties over the system, the use of a single time step can be inefficient, as it may be much smaller than required for low-velocity or low-density regions. This has been addressed by a recent improvement to the DSMC algorithm \cite{Gallis:2009a}, where a local time step was introduced for the collision step. Performing the collision step is computationally expensive, so this improvement can lead to a great increase in the efficiency of calculations.

With the use of LATS, there are two time steps of importance for each cell: (i) The global time step $\tsg$, which is the fundamental increment of time in all cells of the system. 
 The global time after $k$ steps is specified as $t_g=\sum_{i=1}^{k}\tsg$, and during each increment of $\tsg$, collisionless evolution is performed  [i.e., Eqs.~(\ref{collevolv1})$-$(\ref{collevolv3}) with $\ts \to \tsg$].  (ii) The local time step for the cell $\tsc$, which is the desirable time scale for performing collisions in this particular cell. Note $\tsg =\min\{\tsc\}$, i.e., we choose the global time step to be the smallest value of $\tsc$ over all cells in the system at the end of each step.\footnote{If $\tsg$ is sufficiently large that the accuracy of the collisionless evolution is compromised, $\tsg$ is split into smaller increments for this evolution.}

A collision step is performed at the global time step when, at least, a time of $\tsc$ has passed since the last collision step for the cell under consideration [see Fig.~\ref{fig:diagr3}(b)]. To implement this, we introduce a cell timer $t_c$, indicating the time up to which collisions have been accounted for in the cell. In general, $t_c<t_g$ and is incremented by $\tsc$ during each collision step. Performing collisions in this way ensures that $t_c$ is within $\tsc$ of $t_g$ at all times,\footnote{If a cell becomes unpopulated ($N_c=0$), $t_c$ may not have been updated such that $t_c=t_g$ before the test particles leave the cell, which decreases the collision rate. However, $\tsc$ is chosen such that this effect is negligible.} and at the end of the simulation, all $t_c$ are updated to the final time by performing collisions with $\tsc=t_g-t_c$.

In our simulations, $\tsc$ is chosen to be small compared to the relevant collision and transit times of the cell. In detail, these time scales,
\begin{subequations}
\begin{align}
\tau^{\rm{coll}}_c&=\bracb{n_c \overline{v_r \tcs }}^{-1},\\
\tau^{\rm{max}}_c&=\bracd{ n_c \mkrcs}^{-1},\\
\tau^{\rm{tr}}_c&=\min\bracd{\frac{\Delta x_{c} }{ \overline{v}_x},\frac{\Delta y_{c} }{ \overline{v}_y},\frac{\Delta z_{c} }{\overline{v}_z}},
\end{align}
\end{subequations}
are the mean-collision time, the maximum collision time, and the mean-transit times of the cell, respectively.  These expressions are evaluated at the end of each collision step, and the average speeds $\braca{\overline{v}_x,\overline{v}_y,\overline{v}_z }$ are given by averaging over all the test particles within the cell, while $\overline{v_r \tcs}$ is the average of  $v_r \tcs$ over the particles tested for collisions.   The cell widths $\braca{\Delta x_{c},\Delta y_{c},\Delta z_{c}}$ correspond to the cell under consideration [e.g., $\Delta x_{c}$ is the adaptive subcell $x$ width if the LAC is used, and the master bin width ($\Delta x$) otherwise].   

In terms of these time scales, we take 
\begin{equation}
\tsc = \min\bracd{\eta_{\rm{coll}} \tau^{\rm{coll}}_c, \eta_{\rm{max}} \tau^{\rm{max}}_c,\eta_{\rm{tr}} \tau^{\rm{tr}}_c}, \label{eq:dtc}
\end{equation} 
where $\eta_{\rm{coll}}$, $\eta_{\rm{max}}$, and $\eta_{\rm{tr}}$ are constants less than unity. At the end of each collision step, $\tsc$ is reset by \eqr{eq:dtc}. Whenever $\tsc$ is established without performing a collision step, i.e., beginning of the simulation or when the subcells are collapsed or expanded, we take it to be  $\tsc = \min\bracd{\eta_{\rm{max}} \tau^{\rm{max}}_c,\eta_{\rm{tr}} \tau^{\rm{tr}}_c}$. 

For the accurate simulation of dynamics, it is required that $\tsc \ll  \tau^{\rm{coll}}_c$ as well as $\tsc \ll  \tau^{\rm{tr}}_c$. We also require that it is unlikely for a particle to undergo multiple collisions in a collision step (accounted for by $\tau^{\rm{max}}_c$). These requirements are ensured by the constants $\eta_{\rm{coll}}$, $\eta_{\rm{max}}$, and $\eta_{\rm{tr}}$, which are optimized for the desired accuracy.

Care has to be taken when the LATS is implemented in conjunction with the LAC scheme, as the cells can change dynamically during the evolution (cells can be resized, can be added or can be removed). 
Our procedure for dealing with dynamically changing subcells is as follows:
As each master cell is considered in turn, if the number of LAC subcells changes, a new layout of LAC subcells must be established. If the number of these subcells increases, then each of these new cells inherits the $t_c$ of the original cell. Alternatively, if the number of subcells decreases, then the new cells are formed by merging old cells. 
In general, the values of $t_c$ for each of the cells to be merged are different, and we take the new value of $t_c$ to be the largest of these. This requires $t_c$ of the old cells to be updated to the new $t_c$, thus, collision steps are performed within the old cells before merging, using the time difference.

When the LAC scheme is implemented with small threshold numbers (e.g., $\nth < 5$) and the number of test particles is large ($\nt>10^6$), it can become inefficient to implement the LATS in conjunction with the LAC subcells. In such regimes, very dense grids of LAC subcells typically arise, for which the computational intensity of the LATS and memory requirements become too great. Furthermore,  small cell sizes lead to  excessively small time steps (e.g., $\tau^{\rm{tr}}_c$ is proportional to the cell size), which further reduces the algorithm efficiency. In these cases, it is more efficient to implement the LATS for the master cells (i.e., only the master cells have a time counter and desired time step) and implement collisions in all the LAC subcells using that same desired time step.

\section{Tests and optimal parameters}\label{SecTests}

In this section, we develop tests relevant to ultra cold systems that we use to validate and to explore how to optimize the performance of the DSMC algorithm by quantifying the effects of the adaptive enhancements.
Primarily, we are interested in the quality of the representation of the phase-space distribution, since this is of fundamental importance for  accurate Boltzmann evolution. In particular, we address the effects of increasing the number of test particles and refining the grid on collision rates as compared to exact results.

\subsection{Analytic results}\label{sec:analytic}
We  develop  benchmark analytic results to calibrate the algorithm against. 
To do this we consider the equilibrium (Maxwell-Boltzmann) distribution function for a harmonically trapped gas
\begin{equation}
\dfeq  \equiv\np \braca{\beta \hbar \omega}^3 \exp\bracd{-\beta\bracb{\frac{p^2}{2 m} + \pot}}, \label{eq:eqdist}
\end{equation} 
where 
\begin{equation}
\pot= \frac{m}{2}\braca{\omega_x^2x^2+\omega_y^2y^2 +\omega_z^{2}z^{2}} \label{eq:pot}
\end{equation}
is a harmonic trapping potential and  $\omega = \braca{\omega_x\omega_y\omega_z}^{\frac{1}{3}}$.

The total collision rate is given by
\begin{equation}
R= \frac{\sigma_0}{m} \int\frac{d^{3}p}{h^3} \int\frac{d^{3}p_1}{h^3} \int d^{3}r \bracc{\bp_1-\bp} ff_1. \label{eq:colrateint}
\end{equation}
Here, we have taken the differential cross section to be velocity independent  to give a total cross section of $\sigma_0$. 
Evaluating this expression   for the equilibrium cloud, \eqr{eq:eqdist}, we obtain
\begin{equation}
R_{eq}= \frac{\np^2}{2 \pi^2} m \beta \omega^3 \sigma_0.\label{eq:colrate}
\end{equation}

As we are concerned with simulating the collisions of equilibrium clouds, it will be useful to consider the instantaneous distribution,
\begin{equation}
\dfcol= f_{eq}\braca{\bp+ {p}_0\hat{\mathbf{z}},\br}+f_{eq}\braca{\bp- {p}_0\hat{\mathbf{z}},\br}, \label{eq:f_1}
\end{equation}
which corresponds to two spatially overlapping clouds with equilibrium shapes that are traveling with opposite momenta $\pm p_0$ along the $z$ direction. The total collision rate for this case is 
\begin{align}
R_{coll}= & \frac{\np^2}{2\pi^2}m \beta \omega^3 \sigma_0 \left[ 2+\exp\braca{-p_0^2\frac{ \beta}{m}}  \right. \nonumber 
\\  &\left.{}+ \frac{1}{2 p_0}\sqrt{\frac{\pi m}{\beta}}\braca{1+2 p_0^2 \frac{\beta}{m}}\erf{p_0 \sqrt{\frac{\beta}{m}}}\right]. \label{eq:colratecollide}
\end{align}
For small $p_0$, the term in the square brackets scales as $4+\frac{2}{3}\beta p_0^2/m + O\braca{p_0^4}$, showing that, for $p_0=0$, \eqr{eq:colratecollide} reduces to \eqr{eq:colrate} with $\np\to2\np$, as expected. While for large $p_0$, it scales as $2+\sqrt{\beta \pi/m}p_0 + O\braca{p_0^{-1}}$. The first term corresponds to the intra cloud collisions, while the linear term is that of which is obtained for momentum distributions of vanishing width, i.e., Dirac $\delta$ functions $\delta\braca{\bp \pm p_0 \hat{\mathbf{z}}}$.

\subsection{Grid parameters and test-particle number} 
To investigate the accuracy with which collisions are treated, we compare the numerical collision rate to the exact values in \eqrs{eq:colrate} and (\ref{eq:colratecollide}). To do this, we calculate the relative error of the numerical collision rate  and examine its dependence on the number of test particles and grid refinement.\footnote{The relative error in the collision rate is independent of $\np$ and $\sigma_0$.}

\subsubsection{Numerical collision rate} 
For the purpose of comparison, we need to extract a collision rate from the DSMC representation of $\df$. To do this, we evaluate the mean number of collisions in each cell over some time $\tsc$. Hence, in each cell, the mean collision rate is 
\begin{equation}
R_c\approx 2  \alpha \sum_{\braca{ij}}^{\mr} \frac{\pcolr}{\cellv \tsc}, \label{eq:lcrsimulation}
\end{equation}
where $\braca{ij}$ indicates the indices of the $\mr$ selected collision pairs in the cell. Thus, the total collision rate for the system is
\begin{equation}
R=\sum_{\rm{cells}}R_c \cellv. \label{eq:tcrsimulation}
\end{equation}
By calculating the collision rate in this way, we are, in effect, directly performing a Monte Carlo integration for the integral (\ref{eq:colrateint}), which is the basis of the derivation of the collision probability in \refrs{Jackson2002a,Wade:2010a}. The time step for the cell $\tsc$ is somewhat arbitrary, and we choose it to give $\mr = \lfloor {N_c/2}\rfloor $ collision pairs. 

A convenient length scale for the trapped system is given by the thermal widths $W_{x}= \sqrt{ {2k_BT}/{m{\omega_{x}^2}}}$, etc., and we choose the master cell widths such that the resolution in each direction (relative to these widths) are the same, i.e., 
\begin{equation}
\gamma=\frac{\Delta x}{W_x}=\frac{\Delta y}{W_y}=\frac{\Delta z}{W_z}. \label{eq:binparam}
\end{equation}
In what follows, $\gamma$ will serve as an important parameter to specify the fineness of the spatial resolution.

\subsubsection{Accuracy} \label{sec:Accuracy}

To increase the accuracy of our numerical calculation of the total collision rate, we must improve the accuracy of our representation of  continuous distribution $\df$ or take more samples. In DSMC, $\df$ is represented in two ways: (i) the test-particle swarm, (ii) the grid of cells that sample the test-particle swarm. \appr{app:M_c} shows that, without cell adaption [i.e., LAC or LATS], (i) and (ii) are largely decoupled. However, simply decreasing the size of the master cells can cause large statistical fluctuations in the number of collisions, as single occupation of a cell becomes more common, hence, requiring a larger number of samples.

\begin{figure}[h]
\begin{center}
    \includegraphics[width=.9\linewidth]{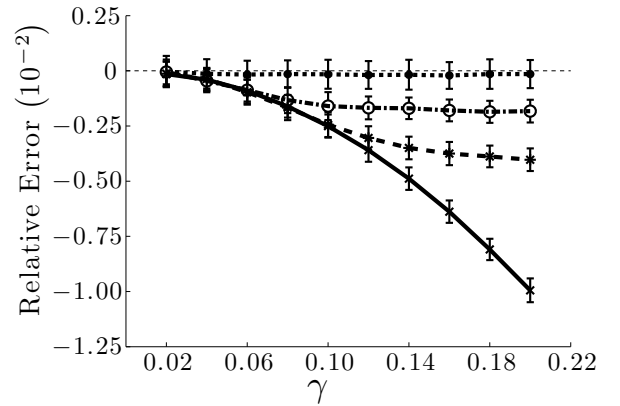}
\caption{The relative error of the total collision rate, \eqr{eq:colrate}, for the equilibrium distribution $\dfeq$ against $\gamma$ with $\nt = 10^7$ is shown for the cases without   (solid line) and with cell adaption where $N_{th}= 2$ (dotted line), $150$ (dashed-dotted line), and $500$ (dashed line). The results shown here are averaged over 200 initial conditions, while the error bars give the standard deviation.
Without adaption, the error increases with increasing $\gamma$, since $\dfeq$ becomes more coarsely grained. However, with the inclusion of adaption, this behavior is combatted as the LAC subcells adapt accordingly. 
We obtain the initial conditions for the test particles from $\dfeq$ using the Monte Carlo acceptance-rejection method.
System parameters: The harmonic potential is chosen to be the same as that used in the ultra cold collider experiment with $\omega_x=\omega_y = 2 \pi \times 155 \, \mathrm{Hz}$ and $\omegaz = 2 \pi \times 12 \, \mathrm{Hz}$,  $\np=2\times10^5$ and $T=600 \, \mathrm{nK}$.
} \label{fig:accuracy}
 \end{center}
\end{figure}

Our LAC scheme essentially establishes a local maximum size of the cells (i.e., maximum error), which is set by $\nt$, $\nth$, and $\denr$. In our results, this is seen for the collision rate of the equilibrium cloud given in  \figr{fig:accuracy}. These results show that the magnitude of the relative error does not continue to increase with increasing $\gamma$ (as it does in the unadapted case) but tends to a constant dependent on $\nth$. With decreasing $\nth$, smaller cell sizes are achieved, hence, lower error.\footnote{Care needs to be taken with other adaptive schemes, since the approach outlined in \appr{app:M_c}, to remove statistical biasing, neglects to take into account statistical fluctuations from other sources (e.g. volume), which may become important \cite{Garcia2009a}.} However, we restrict ourselves to $\nth \geq 2$ to avoid the increasingly large statistical fluctuations mentioned earlier.
The results in \figr{fig:accuracy} remain qualitatively similar for different values of $\nt$, however, the fluctuations (i.e., error bars \figr{fig:accuracy}) increase with decreasing test-particle number.

It is worth noting that systems with identical density distributions are coarse grained in the same fashion (provided $\nt$ is the same when using the LAC scheme), hence, they have the same accuracy. For example, the equilibrium (\ref{eq:eqdist}) and collision (\ref{eq:f_1}) distributions have identical relative error profiles as seen in \figr{fig:accuracy}. However, if a system is dynamically changing and no adaption was employed, evolving to a more dilute system would decrease the magnitude of the relative error, while increasing if becoming denser. For adaptive schemes, this is not an issue, as the cell sizes automatically adjust to this change.

\subsubsection{Performance considerations}

The results in Fig.~\ref{fig:accuracy} show that the following cases approximately have the same relative error in collision rate:
[{\bf SIM1}] an unadaptive  simulation with $\gamma=0.02$,  [{\bf SIM2}] a LAC simulation with $N_{th}=2$ and  $\gamma=0.2$ (we also include the LATS for dynamics in SIM2).\\
 A  fuller picture of the merits of using either of these approaches for a simulation requires us to understand their resource requirements.  
 
\noindent {\bf Speed:}  We find that, with our code SIM2 is approximately five times faster than SIM1 for near-equilibrium evolution. Note, we only use the LATS scheme in SIM2 for the master cells (as discussed at the end of Sec.~\ref{sec:LATSs}). It should also be noted that this performance  indicator is dependent on the code implementation and physical problem under consideration (i.e., equilibrium cloud versus highly nonequilibrium situation).  

\noindent {\bf Storage:} 
 SIM1 requires $\sim 5 \times 10^7$ master cells, while SIM2 requires $\sim 5 \times 10^4$ master cells with a maximum of $4096$ LAC subcells within a master cell (typically requiring a total of $\sim7\times10^6$ LAC subcells).

\subsection{Collisions between clouds:  Comparison to simple methods} \label{sec:comparison}

\begin{figure}[t]
\begin{center}
    \includegraphics[width=.9\linewidth]{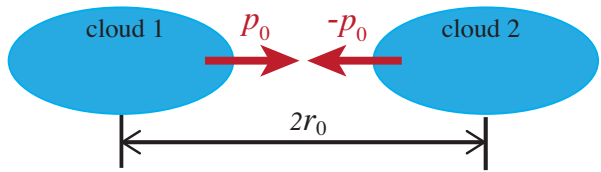}
\caption{(Color online) Schematic of the ultra cold collider  used in Sec.~\ref{sec:comparison}. Two clouds initially separated by a distance of $2r_0$ collide at a relative momentum of $2p_0$. The number of atoms that have scattered out of the clouds, after they have passed through each other, is referred to as $\nsc$.
} \label{fig:uccollider}
 \end{center}
\end{figure}
In this subsection, we consider the collision of two equilibrium clouds in a harmonic trap, $f_{eq}\braca{\bp \pm p_0\hat{\mathbf{z}},\br \mp r_0 \hat{\mathbf{z}}}$, shown schematically in Fig.~\ref{fig:uccollider}. We study this collision using our DSMC algorithm and compare its results to a simplified model that has been used previously to analyze this problem. Initially, the two clouds are centered at locations separated by a distance of $2r_0$ along the $z$ direction, chosen to ensure that (initially) the clouds do not overlap. 
The clouds approach each other, moving at a relative momentum of $2p_0$, and when they overlap, collisions scatter atoms out of the clouds. 
Here, our main interest is the total number of such scattered atoms $\nsc$, after the two clouds have completed passing through each other.
 
The  simple model we consider was used in Ref.~\cite{Thomas:2004a} (see also, \refr{Band:2000a}) and was derived from the Boltzmann equation  description of the colliding clouds by making the following approximations: {\bf (a1)} the harmonic potential is ignored (collision taken to be in free space); {\bf (a2)} the momentum distribution of each cloud is replaced by $\delta\braca{\bp \pm p_0 \hat{\mathbf{z}}}$; {\bf (a3)} the dynamics of scattered atoms are neglected. These approximations lead to an equation for the densities $n_i$ of cloud $i=1,2$  of
\begin{equation}
\braca{\frac{\partial}{\partial t}\pm\frac{v_r}{2}\frac{\partial}{\partial z}}n_{i}\braca{\br,t}=-v_r\sigma_0 n_{1}\braca{\br,t} n_{2}\braca{\br,t},\label{eq:simplem}
\end{equation}
where $v_r=2p_0/m$. We can solve these equations directly using a pseudo-spectral method. 

An analytic expression may be derived with an additional approximation: {\bf (a4)} The loss of atoms is small enough such that the shape of the densities do not deform but remain Gaussian while the normalization of each cloud $\np$ decreases. Using this, one can integrate \eqr{eq:simplem} over all position space to find the total number of scattered atoms from the collision,
\begin{equation}
\nsc = \frac{\np^2}{4\pi} m \beta \omega_x\omega_y \sigma_0.\label{eq:simplem2}
\end{equation}

Following the terminology established in experiments, we characterize the collider kinetic energy in temperature units by the parameter $\tcoll\equiv \mu v_r^2/2k_B$, where $\mu=m/2$ is the reduced mass. 
As shown in \secr{sec:analytic}, when considering the limiting behavior of \eqr{eq:colratecollide}, the approximation (a2) is satisfied when $\tcoll\gg  T$ (which is the case for collider velocities we consider here). That the momentum distributions can be replaced with Dirac $\delta$ functions is consistent with many-body quantum statistics not playing a significant role in the scattering that occurs when the two clouds collide. However, the internal motion of each cloud can be influenced by quantum statistics.

As the full DSMC solution includes the dynamics of scattered atoms, it is useful to split the scattered atoms into two groups: (i) scattered atoms that have not undergone any subsequent collisions, (ii) scattered atoms that have undergone additional collisions, including all collision partners.\footnote{We include atoms that are scattered out of cloud 1 or 2 by a collision with an already scattered atom.} All of the scattered atoms predicted by \eqrs{eq:simplem} and (\ref{eq:simplem2}) are of group (i).

In \eqrs{eq:simplem} and (\ref{eq:simplem2}), $\nsc$ is independent of the details of the differential cross section (only depending on the total cross section), and this is largely true for the full solution in the case considered here. Thus, it is convenient to take $\sigma_0=8 \pi a_{sc}^2$, which is of the form of the total cross section for \textit{s}-wave scattering in the low-collision energy limit with scattering length $a_{sc}$. Additionally, $\nsc$ in both equations is independent of $v_r$, i.e., $\tcoll$. However, this is not the case for the full solution, since the collision occurs in a trap. For example,  if the radial confinement is tight, then a scattered atom can oscillate out and back in the  radial plane and can recollide  (depending on the timescale over which the collision proceeds). 
Here, we choose to operate in a regime where these effects are small and the simple model should accurately describe the full solution. 
To do this, we choose parameters such that  $\tcoll = 300 \, \mathrm{\mu K}$,\footnote{For the full DSMC solution, the clouds accelerate as they approach the trap center, and we take the value of $p_0$ that they obtain at the trap center as the value to compare against the simple model.} giving a short time scale for the collision.

\begin{figure}[h]
\begin{center}
    \includegraphics[width=.9\linewidth]{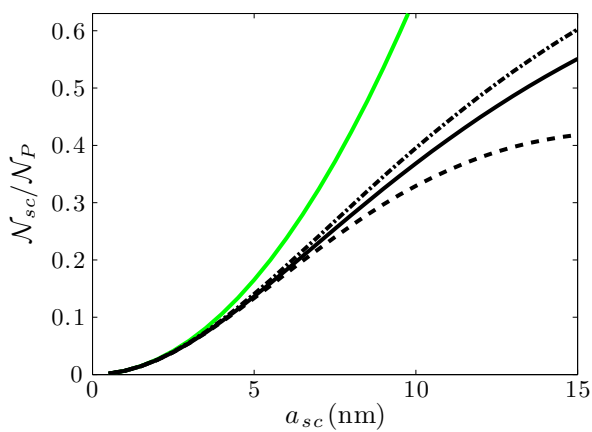}
\caption{(Color online) The fraction of scattered atoms due to the collision of two equilibrium clouds as a function of $a_{sc}$. Equation (\ref{eq:simplem2}) [solid green (gray) line] has poor agreement with the solution of \eqr{eq:simplem} (solid black line) for $\nsc/\np > 0.05$, since approximation (a4) is no longer valid. Group (i) scattered atoms (dashed black line)  and total scattered atoms [groups (i) and (ii)] (dashed-dotted black line) from the DSMC solution.
The system parameters are given in \figr{fig:accuracy}, and for the DSMC simulation, $\gamma=0.2$, $\nt= 10^7$, and $\nth=2$. The standard deviation error is not shown as it is on the order of the line width.} \label{fig:models}
 \end{center}
\end{figure}

The results of \eqrs{eq:simplem} and (\ref{eq:simplem2}), as well as the full solution, are shown in \figr{fig:models} for varying $a_{sc}$. All models agree well in the low scattering regime $\nsc/\np < 0.05$, while for higher scattering fractions, approximation (a4) becomes invalid, and the dynamics of the scattered atoms becomes increasingly important. However, the solution of \eqr{eq:simplem} agrees to within $10\%$ of the relative error to the total [groups (i) and (ii) combined] scattered fraction given by the full solution over the whole range. 

We can modify the collision problem and DSMC method to a regime that is exactly described by the simplified equation (\ref{eq:simplem}). To do this, all particles are taken to have momentum $\pm p_0$ along the $z$ axis (the components of momenta   in the $xy$ plane are zero) and evolve without an external trapping potential. Consistent with the approximations going into Eq.~(\ref{eq:simplem}), whenever a pair of particles undergoes a collision, it is removed from the system (eliminating any need for consideration of multiple collisions). Due to the form of the distribution function, nearest-neighbor collisions cannot be used.\footnote{Particles have no transverse momenta, thus particles from the same cloud never leave the proximity of each other. Hence, it is required that a particle from one cloud is closest to a particle from the other cloud before a collision can occur, which results in a decreased number of collisions.}    

\begin{figure}[h]
\begin{center}
    \includegraphics[width=.9\linewidth]{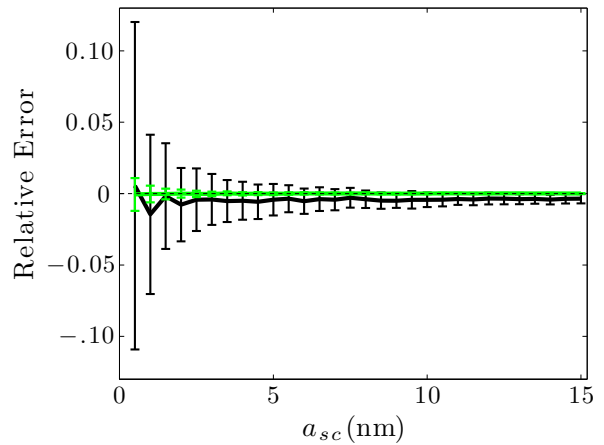}
\caption{(Color online) The relative error of $\nsc$ as calculated by the DSMC solution of \eqr{eq:simplem} [see text] to that of our pseudo spectral solution of Eq.~(\ref{eq:simplem}). Here, we show the two cases $\nt=10^5$ (black) and $10^7$ [green (gray)] with $\gamma=0.2$, $\nth=25$, and the system parameters given in \figr{fig:accuracy}. We use $\eta_{\rm{coll}}=0.01$, $\eta_{\rm{max}}=0.1$, and $\eta_{\rm{tr}}=0.01$. 
The results shown here are averaged over 200 simulations, while the error bars give the standard deviation.
} \label{fig:matchingerror}
 \end{center}
\end{figure}

The relative error of $\nsc$ as calculated by the DSMC solution to that of our numerical solution of Eq.~(\ref{eq:simplem}) is shown in \figr{fig:matchingerror} for the two cases $\nt=10^5$ and $10^7$. The excellent agreement of the two results is a good test that the DSMC method is correctly implemented. The error bars represent the statistical fluctuations of the DSMC results. These fluctuations reduce with increasing $a_{sc}$ as $\nsc$ increases, while between the two cases, they are reduced by a factor of $10$, since they also decrease  with increasing $\nt$ (to be definite, these fluctuations are given by the inverse square root of the number of scattered test particles).

\section{Many-body simulation of an ultra cold collider}\label{SecCollider}

In this section, we demonstrate the application of our DSMC algorithm to the simulation of the ultra-cold-atom collider reported in Ref.~\cite{Thomas:2004a}. The main extension,  over the DSMC collision test presented in Sec.~\ref{sec:comparison}, is the inclusion of the full two-body collisional cross section needed for a realistic microscopic description of the collisional interactions. We then extend our consideration to the long-time dynamics of the collider and how the system progresses to equilibrium.

\subsection{Collisional cross section}

Experiments realizing the ultra-cold-atom collider were conducted with $^{87}$Rb, which is bosonic, prepared in a single hyperfine spin state ($F=2,m_F=2$).
The  wave function for two such colliding atoms is required to be symmetric, hence, only the even partial-wave terms in \eqr{eq:dcs} contribute to the differential cross section. At the collision energies of the experiment, only the first two even terms contribute, $l=0$ and $l=2$ (\textit{s} and \textit{d} waves). Thus, the differential cross section reduces to
\begin{align}
\dcs=& \frac{4 \hbar^2}{m^2 v_r^2} \left[ \overbrace{4\sin^{2}\phasea}^{\text{\textit{s} wave}} + \overbrace{25\sin^{2}\phaseb\ucos^{2}}^{\text{\textit{d} wave}}\right.\nonumber \\
 &\left.{}+\underbrace{20\cos\braca{\phasea-\phaseb} \sin\phasea \sin\phaseb\ucos}_{\text{\textit{s}- and \textit{d}-wave interference}} \right].
\end{align} 
Taking care to integrate over only half the total solid angle to avoid double counting, the total cross section $\tcs$ is given by the sum of the total \textit{s}- and \textit{d}-wave cross sections,
\begin{equation}
\tcs=\frac{32\pi\hbar^{2}}{m^2 v_r^2}\braca{\sin^{2}\phasea + 5\sin^{2}\phaseb }. \label{eq:tcs}
\end{equation}

\begin{figure}[h]
\begin{center}
    \includegraphics[width=.9\linewidth]{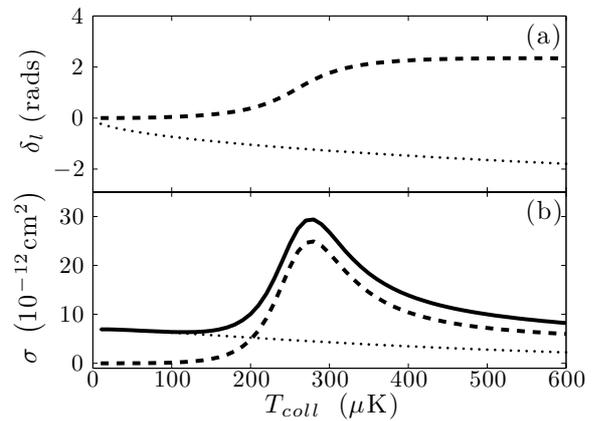}
\caption{(a) Numerically calculated \textit{s}-wave (dotted line) and \textit{d}-wave (dashed line) phase shifts of \refr{Thomas:2004a}. (b) \textit{s}-wave (dotted line), \textit{d}-wave (dashed line), and total (solid line) cross sections. \label{fig:phaseshifts}}
 \end{center}
\end{figure}

Calculation of the collision energy dependence of the phase shifts $\phasea$ and $\phaseb$ is a nontrivial task. The values that we use in our simulations [\figr{fig:phaseshifts}(a)] are those calculated by Thomas \etal and reported in Ref.~\cite{Thomas:2004a}.
Over the range of collision energies shown in \figr{fig:phaseshifts} the interference between  \textit{s}- and \textit{d}-wave scatterings can be important, and  a \textit{d}-wave resonance also occurs. The \textit{d}-wave resonance can be seen in \figr{fig:phaseshifts}(b) by the peak of the total cross section, attributed to the large \textit{d}-wave cross section.

\subsection{DSMC simulations}

\begin{figure}[h]
\begin{center}
    {\includegraphics[width=\linewidth]{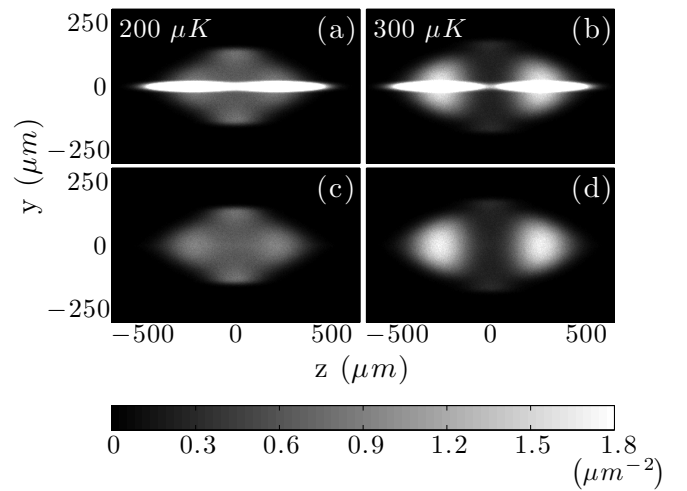}}
\caption{Column densities at time $\pi/2 \omega_x$ after the clouds reach the center of the trap. $\tcoll=200\,  \mathrm{\mu K}$ [(a) and (c)] is a regime of \textit{s}- and \textit{d}-wave interference, while $\tcoll=300 \,\mathrm{\mu K}$ [(b) and (d)] is a \textit{d}-wave regime. (c) and (d) only show the scattered atoms. The initial conditions for the clouds are chosen as in \secr{sec:comparison}, and the system parameters are given in \figr{fig:accuracy}, while the simulation parameters are $\gamma=0.2$, $\nt=10^5$, and $\nth=2$. The results were averaged over 200 simulations. Note, we have compared these results to simulations with $\nt=10^7$ also averaged over 200 runs, and we find that the number of scattered particles and the angular scattering distributions agreed to within 1\%.} \label{fig:sdscattering}
 \end{center}
\end{figure}

Using the full energy and angular-dependent-scattering cross section, our DSMC method can provide an \textit{ab initio} prediction for the collider experiments. 
The full and detailed comparison with experiments and what information this reveals about the two-body collisions are beyond the scope of this paper and will be presented elsewhere (although we note that the density images shown in Fig.~\ref{fig:colliderimage} confirm that our approach provides a visually good match to the experimental results).

Here, we present the results of  column densities calculated after two equilibrium  clouds $f_{eq}\braca{\bp \pm p_0\hat{\mathbf{z}},\br \mp r_0 \hat{\mathbf{z}}}$ have collided for the cases $\tcoll=200\, \mathrm{\mu K}$ and $300 \, \mathrm{\mu K}$. Following the experimental procedure \cite{Thomas:2004a}, we calculate these column densities at a  quarter of the radial trap period ($\pi/2 \omega_x$) after the clouds reach the center of the trap. At this time the bulk of the scattered atoms reach their maximal extent in the radial direction. Figures \ref{fig:sdscattering}(a) and \ref{fig:sdscattering}(c) show a regime of \textit{s}- and \textit{d}-wave interference ($\tcoll=200 \, \mathrm{\mu K}$), while \figsr{fig:sdscattering}(b) and  \ref{fig:sdscattering}(d) show a \textit{d}-wave regime ($\tcoll=300 \, \mathrm{\mu K}$). Clearly, the distribution of scattered atoms deviates from the typical \textit{s}-wave halo (e.g., see \refr{Chikkatur2000a}). 

\subsection{Long-time dynamics: Rethermalization}

The idea of using rethermalization of colliding condensates to perform calorimetry has been proposed in \refr{Blakie2007d}, however, no direct simulations were made of the thermalization dynamics.  More generally, there has been significant recent interest in how a quantum system rethermalizes \cite{Rigol2008a}, particularly in systems that might be experimentally realized with ultra-cold-atomic gases (e.g., see \refrs{Rigol2009a,Cassidy2011a}). To date, much of the attention has been focused on integrable or nearly integrable systems where numerical solutions are available for small samples of atoms (typically $\np<10^2$). However, in such regimes, thermalization is often inhibited or strongly is effected by constraints (e.g., see \refr{Kinoshita2006a}) as well as being difficult to explore experimentally due to the small atom number  (or requiring many similarly prepared systems to get a good signal).

\begin{figure}[h]
\begin{center}
    \includegraphics[width=\linewidth]{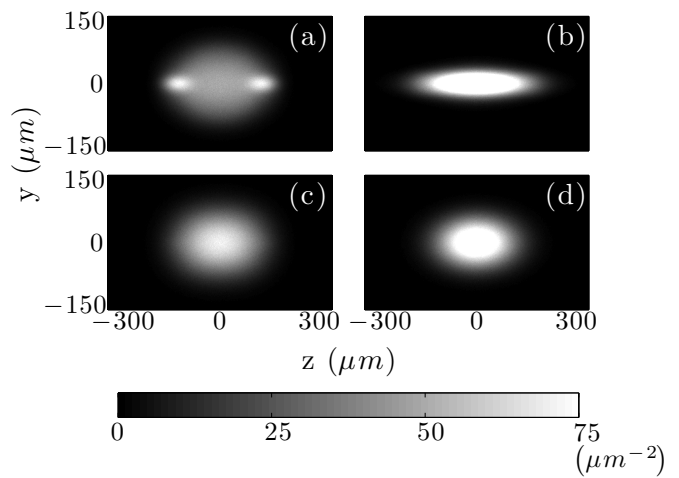}\\ 
    \caption{Column densities illustrating the long time dynamics of rethermalization. (a) At time $t \omega_z = 7.04$ after the clouds have passed through each other twice. The colliding clouds are still visible (density peaks). When the colliding clouds are depleted, the system continues to evolve through collective oscillations  that are illustrated by the images (b) and (c) at  $t \omega_z = 18.85$ and $19.60$, respectively. The decay of these collective oscillations occurs on a slower time scale than the depletion of the colliding clouds, and the distribution does not take on the equilibrium distribution until much later times as seen in (d) at  $t \omega_z = 500.02$. The trap frequencies are $\omegaz = 2 \pi \times 50 \, \mathrm{Hz}$ and $\omega_x=\omega_y = 2 \omega_z$, and each of the initial clouds has $\np= 10^6$ and $T=600 \, \mathrm{nK}$. We use an isotropic differential cross section with $a_{sc}=10$ nm. The initial separation is chosen such that there is insignificant overlap of the clouds. The momenta are chosen to give $\tcoll=32.4 \,\mathrm{\mu K}$, giving a final equilibrium temperature of $T=6 \, \mathrm{\mu K}$. Note, for an isotropic trap, the system does not completely thermalize without mean-field effects, since the breathing mode does not damp \cite{Guery-Odelin1999a}.} \label{fig:ltd1a}
\end{center}
\end{figure}

\begin{figure}[h]
\begin{center} 
    \includegraphics[scale=0.5]{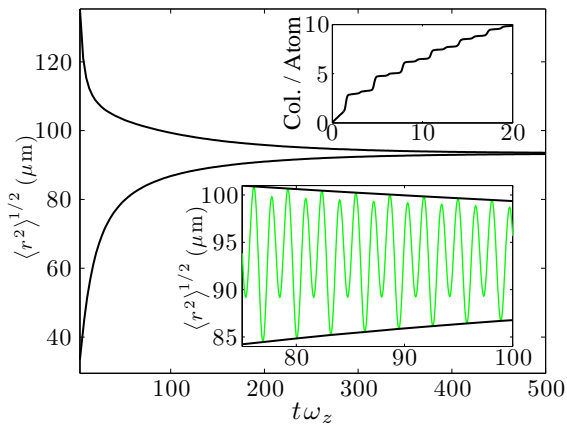}
    \caption{(Color online) Envelope of the oscillations (as seen in the lower inset) of the root-mean-square of $r$. The rapid decay of the envelope within the first ten trap cycles is attributed to the depletion of the colliding clouds, while the slower decay is the decay of the collective modes. The upper inset shows the mean number of collisions per atom. 
    } \label{fig:ltd1b}
\end{center}
\end{figure}

Thus, we are motivated to apply the DSMC method to model the dynamics of colliding ultra cold clouds well past the first collision. As the collisions occur in the trap, the clouds will oscillate back and forth,  recolliding each time, and thus, are provided with the opportunity to rethermalize. This system is much larger and classical in nature than the small quantum systems generically considered for thermalization studies. However, we believe this is an interesting system: first as a bridge between quantum and classical thermalization in ultra cold gases that is practical for experimental investigation. Second, such a system might provide a unique opportunity to test the Boltzmann equation in a regime where the microscopic parameters are precisely known and with well-characterized far-from-equilibrium initial conditions. Few equations in theoretical physics have evoked as much discussion and controversy as the Boltzmann equation -- particularly in reference to the introduction of irreversibility -- such a test could be of broad interest and shed light on some long-standing issues in statistical mechanics.

Our first evidence for thermalization comes from examining the density profiles of the colliding clouds at times after the first collision. Some examples of these density profiles  are shown in Figs.~\ref{fig:ltd1a}(a)-\ref{fig:ltd1a}(d) and reveal that, as time passes, the number of atoms participating in the parametric oscillation of the mother clouds along the $z$ axis decreases as the collisions convert the system to a more isotropic form. Indeed, the system clearly appears to increase entropy and approaches an equilibrium like configuration. 

In order to quantify the approach to equilibrium, it is useful to consider how various moments of the system evolve dynamically.  In Fig.~\ref{fig:ltd1b}, we show the envelope of the oscillations in the position spread moment $\langle r^2\rangle^{1/2}=\langle  x^2+y^2+z^2\rangle^{1/2}$, characterizing the root-mean square of the distance of the particles from the trap center. [Note the oscillations of this moment occur on a much faster timescale and are shown in an inset to Fig.~\ref{fig:ltd1b}.] These results show that the system rethermalizes quite rapidly over the first approximately five trap periods. The number of collisions per particle over the first approximately three trap periods is shown in the inset to Fig.~\ref{fig:ltd1b}. The steps in collision number, which are initially apparent, arise from the periodic recolliding of the clouds. However, as the system is distributed over modes, these steps smooth out. These results show that, during this initial rapid phase of rethermalization, atoms experience $\gtrsim10$ collisions, much greater than the value of $2.7$ often quoted in the literature from the study of 
Wu and foot \cite{Wu1996a}.

After this rapid thermalization phase, the relaxation to equilibrium proceeds more slowly as energy contained within a few low-frequency collective modes waits to be damped.  
 We find that two modes are dominant on long time scales. Most importantly, a mode that oscillates at $2\omega_x (=2\omega_y)$ is dominated by radial breathing character and is well described (both frequency and damping) by the analytic predictions given in Ref.~\cite{Guery-Odelin1999a}.  Also, we note that the rate of relaxation is strongly dependent on the trapping geometry and collision rate.

\begin{figure}[h]
\begin{center}
     \includegraphics[width=0.9\linewidth]{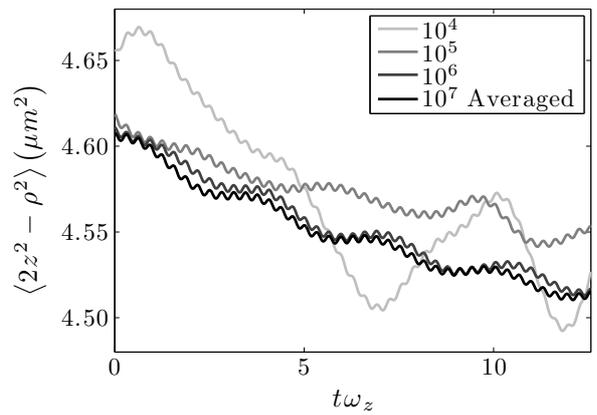} 
\caption{Collective oscillation induced by a small contraction of the radial trap confinement  for a system with $\omegaz = 2 \pi \times 50 \, \mathrm{Hz}$, $\omega_x=\omega_y = 10 \omega_z$,  $\np= 10^4$, and $T=600 \, \mathrm{nK}$. The legend gives $\nt$ used, and as this increases, the results for a single run become increasingly indistinguishable from the ensemble-averaged result for $\nt=10^7$. } \label{fig:ltd2}
 \end{center}
\end{figure}

In relation to thermalization dynamics, it is interesting to revisit the role of test particles in the DSMC simulation. In general, increasing $\nt$ has the effect of reducing fluctuations in a simulation and, hence, the number of trajectories needed to obtain an ensemble average. However, in order to gain a better understanding of typical results (and, hence, fluctuations) that might be expected in experiments, it is necessary to take $\nt=\np$. To illustrate this, we show some results for a small amplitude collective-mode oscillation in Fig.~\ref{fig:ltd2} for a system with $\np=10^4$ and various numbers of test particles. As the number of test particles increases, the results become increasingly indistinguishable from the ensemble-averaged results. However, for  $\nt=\np$, the individual trajectory deviates significantly.  

We emphasize that our simulations for thermalization in this section have been performed for the case of purely \textit{s}-wave scattering. A detailed study of thermalization, including higher-order partial waves (e.g., as the collision energy is scanned across the \textit{d}-wave resonance), would be needed for detailed comparison with experiments in this area but is beyond the scope of this paper. Along these lines, we would like to note an interesting interplay between the partial waves that has been shown in the study of the thermalization of mixtures by Anderlini and Gu\'{e}ry-Odelin \cite{Anderlini2006a}. In that paper, they performed an analytical study of near-equilibrium thermalization of a two-component mixture and showed that the thermalization time (unlike the collision rate) depended on the interference between the scattering partial waves.

\section{Conclusions and outlook}
In this paper, we have presented a DSMC method for simulating the dynamics of nondegenerate ultra cold gases. The motivation for our paper was experiments in which two clouds were collided at high relative velocity. In order to simulate this highly nonequilibrium regime, we have adopted several modern enhancements of the DSMC algorithm (i.e., locally adaptive time steps and nearest-neighbor collisions, introduced in other fields) but not previously used for cold-atom simulation. We have verified that our algorithm is accurate by comparison to a range of analytic results and simplified models.  We have also provided some benchmarks of the performance of our algorithm against traditional DSMC to quantify the computational efficiency.

In order to quantitatively describe the collision experiments, we have included the full energy dependence of the $s$- and $d$-wave scatterings in the differential cross section. We have presented examples of the scattered distributions for the regime of experiments revealing the $d$-wave shape resonance. We have also considered the long-time dynamics of the colliding clouds, allowing them to recollide many times in the trap, observing how they approach equilibrium. Our paper suggests that this might be a fruitful system for future experimental studies to test the accuracy of the Boltzmann equation and to better understand thermalization.

A future application of this paper will be to produce a complete dynamical finite temperature theory. Using a simple DSMC algorithm, Jackson and co-workers \cite{Jackson2001a,Jackson2001b,Jackson2002a,Jackson2002b,Jackson2002c} have already implemented the ZNG formalism \cite{Zaremba1999a}. In the future, we intend to perform a similar extension to c-field formalism \cite{cfieldRev2008}. Having  efficient procedures for evolving the c-field equations that describe the low-energy condensed or partially condensed part of the system \cite{Blakie2005a,Blakie2008a}, the algorithm described in this paper provides the basis for an efficient scheme for simulating the high-energy incoherent modes.  Another avenue of investigation that we are currently exploring is an efficient and accurate way to simulate the quantum Boltzmann equation. That is, to include the effects of Bose-stimulated or Pauli-blocked collisions.
\section*{Acknowledgments}
We acknowledge fruitful discussions with A.~S.~Bradley.   This work was supported by Marsden Contract No. UOO0924 and FRST Contract No. NERF-UOOX0703. 
 
 \appendix
 
\section{Number of tested collisions} \label{app:M_c}
 
 The Boltzmann equation describes the evolution of the continuous distribution $\df$. However, the replacement of $\df$ with a swarm of test particles introduces fluctuations, which do not correspond to physical fluctuations when $\np\neq\nt$. As a result, hydrodynamic quantities are required to be obtained from the averages of mechanical variables, not the average of their instantaneous values \cite{Garcia2006a}. 
 
\begin{figure}[t]
\begin{center}
    \includegraphics[width=\linewidth]{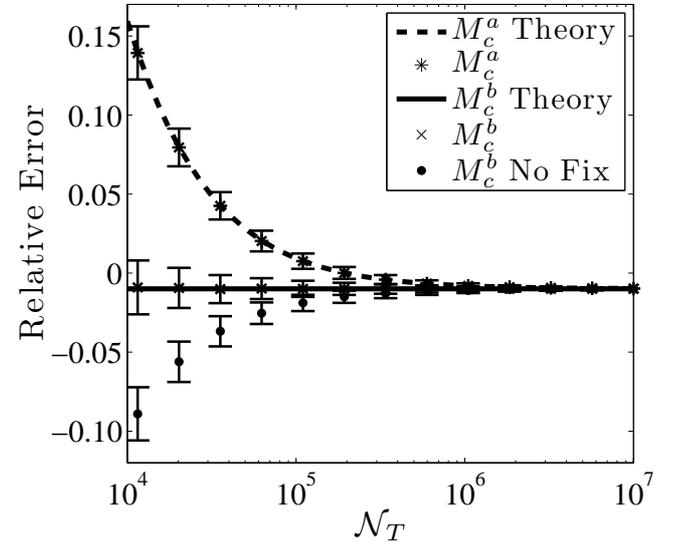}
\caption{Relative error of the numerical total collision rate in the case of the equilibrium distribution (\ref{eq:eqdist}) for the choices of $M_c$, where the error bars indicate the standard deviations of 500 averages. Here, the original DSMC algorithm is implemented with $\nt=10^7$ and bin parameter (\ref{eq:binparam}) $\gamma=0.2$. The choice $M_c^a$ is seen to diverge for low $\nt$ as $\overline{N_c}$ in \eqr{eq:badav} becomes appreciable, which agrees well with the theoretically calculated error ($M^a_c$ theory) using \eqr{eq:badav}.
Using $M_c^b$ removes this divergence, and the error is seen to agree well with the expected error for this discretization ($M^b_c$ theory). 
The final data set ($M^b_c$ no fix) demonstrates the error that arises when $\mr$ having non integer values after rescaling, is not accounted for (see \secr{sec:colprob})  [i.e., not including the ceiling function in Eq.~(\ref{eq:numcol})]. The system parameters are given in \figr{fig:accuracy}. } \label{fig:derror}
 \end{center}
\end{figure}

 In these stochastic particle methods, the collisions of test particles inherently average the instantaneous values of the collision rate. This leads to a biasing of the total collision rate when cells have low occupation numbers (see \figr{fig:derror}). 
To a good approximation \cite{Hadjiconstantinou2003a}, the probability of $N_c$ test particles within a cell is given by the Poisson distribution of which the variance is equal to the mean, i.e., $\overline{\delta N_c^2} = \overline{N_c^2} - \overline{N_c}^2 = \overline{N_c}$, where $\delta N_c=N_c - \overline{N_c}$. 
Note that, formally, the correct number of collisions to test (given by elementary scattering theory and the derivation of the collision probability from the collision integral (\ref{eq:icol}) via the Monte Carlo integration \cite{Jackson2002a,Wade:2010a}) is
 \begin{equation}
M^a_c=\frac{N_c^2}{2}.\label{eq:mca}
\end{equation}
However, with  Poissonian fluctuations in $N_c$, we get that the mean collision rate is  $R\propto\overline{M_c}\sim \overline{N_c}^2+ \overline{N_c}$ (but should be $\propto\overline{N_c}^2$). Thus, Poissonian fluctuations can become important when the number of test particles  per cell is small. However, the effect of fluctuations from the finite-test particle number can be bypassed  (e.g., see \refr{Bird2007a}) by instead using the number of possible pairs of test particles,
\begin{equation}
M^b_c=\frac{N_c\braca{N_c - 1}}{2}, \label{eq:mcb} 
\end{equation}
which we have employed in this paper.

To understand the difference in detail, we note that the average calculated by the DSMC simulation (denoted by the asterisks) for \eqr{eq:mca} is 
\begin{equation}
\overline{N_c^2}^*=\overline{N_c}^2 + \overline{N_c} - P_1, \label{eq:badav}
\end{equation}
where $P_1$ is the probability of $N_c=1$ (as the simulation ignores cells with $N_c=1$, for which no collisions occur, and this must be subtracted from the average). While, for expression (\ref{eq:mcb}),
\begin{equation}
\overline{N_c \braca{N_c - 1}}^*=\overline{N_c}^2,
\end{equation}
which gives the correct total collision rate for the physical system as seen in \figr{fig:derror}.

\bibliographystyle{apsrev4-1}


%

\end{document}